\documentclass[useAMS,referee,usenatbib]{biom}
\usepackage{natbib}
\usepackage{amsmath}
\usepackage{booktabs}
\usepackage{multirow}
\usepackage{array}
\usepackage{rotating}
\usepackage{graphicx}
\usepackage{subfig}
\usepackage{enumitem}
\setlist{nolistsep}
\usepackage{xr}
\def\bSig\mathbf{\Sigma}

\title[]{On the Interplay Between Exposure Misclassification and Informative Cluster Size}

\author{Glen McGee$^{1,*}$\email{glenmcgee@g.harvard.edu}, Marianthi-Anna Kioumourtzoglou$^{2}$, Marc G. Weisskopf$^{3,4}$,\\ {\bf Sebastien Haneuse$^{\mathbf 1}$, and Brent A. Coull$^{\mathbf 1,\mathbf 3}$} \\ 
	{ $^{1}$Department of Biostatistics, Harvard T. H. Chan School of Public Health, Boston, MA, U. S. A} \\
	$^{2}$Department of Environmental Health Sciences, Columbia University Mailman School of Public Health, \\ New York, NY, U.S.A. \\ %
	{ $^{3}$Department of Environmental Health, Harvard T.H. Chan School of Public Health, Boston,  MA, U.S.A.} \\
	{ $^{4}$Department of Epidemiology, Harvard T. H. Chan School of Public Health, Boston, MA, U.S.A.}  }

\begin{document}
	
	
	
	
	
	\pagerange{\pageref{firstpage}--\pageref{lastpage}} 
	\volume{000}
	\pubyear{0000}
	\artmonth{000}
	
	
	\doi{000}
	
	
	\label{firstpage}
	
	
	\begin{abstract}
		In this paper we study the impact of exposure misclassification when cluster size is potentially informative (i.e., related to outcomes) and when misclassification is differential by cluster size. 
		First, we show that misclassification in an exposure related to cluster size can induce informativeness when cluster size would otherwise be non-informative. Second, we show that misclassification that is differential by informative cluster size can not only attenuate estimates of exposure effects but even inflate or reverse the sign of estimates. To correct for bias in estimating marginal parameters, we propose two frameworks: (i) an observed likelihood approach for joint marginalized models of cluster size and outcomes and (ii) an expected estimating equations approach. Although we focus on estimating marginal parameters, a corollary is that the observed likelihood approach permits valid inference for conditional parameters as well. Using data from the Nurses Health Study II, we compare the results of the proposed correction methods when applied to motivating data on the multigenerational effect of in-utero diethylstilbestrol exposure on attention-deficit/hyperactivity disorder in 106,198 children of 47,450 nurses.
	\end{abstract}

	%
	
	
	
	\maketitle
	
	
	%

	\section{Introduction}
	\label{s:intro3}
	Environmental exposures such as endocrine disruptors pose a distinct risk to population health because of their potential to affect future generations via epigenetic mechanisms. Multigenerational studies, such as a recent study of diethylstibelstrol (DES) exposure and third generation attention-deficit/hyperactivity disorder (ADHD) \citep{kioumourtzoglou2018association}, investigate the effects of such exposures on outcomes in later generations. These studies, however, are susceptible to two statistical obstacles. First, family size is often related to health outcomes among children---that is, cluster size may be \emph{informative}. Second, exposures are often assessed retrospectively---and hence subject to misclassification.
	
	Cluster size is informative if it is related to the outcome of interest, given covariates---for example if a shared genetic frailty or toxicity affects both fertility and child development. In such a case, marginal (or population-average) inference can proceed via joint marginalized models for cluster size and outcome \citep{seaman2014review,mcgee2019informatively} or weighted estimating equations \citep{williamson2003marginal,benhin2005mean,seaman2014methods}. 
	Cluster-specific inference proceeds via conditionally specified joint models \citep{dunson2003bayesian,gueorguieva2005comments} or generalized linear mixed models (GLMMs) \citep{neuhaus2011estimation}.
	
	Exposure misclassification and measurement error in clustered-correlated settings have a rich history in the statistical literature \citep{carroll2006measurement,yi2016statistical}. 
	Numerous extensions to generalized estimating equations (GEE) have been proposed to permit marginal inference in the presence of mismeasured covariates, including expected estimating equations \citep{wang2000expected,wang2008expected}, corrected estimating equations \citep{yi2005robust,yi2012functional,chen2014marginal}, regression calibration \citep{sanchez2009estimating} and simulation-extrapolation \citep{yi2008simulation}. 
	Methods for GLMMs have similarly been proposed \citep{carroll1997generalized,wang1998bias,lin1999simex,wang1999bias,liang2009generalized,yi2011simultaneous}.
	
	The recent study of DES and ADHD in the Nurses Health Study II \citep{kioumourtzoglou2018association} exhibited signs of both informative cluster size---ADHD was most prevalent in small families---and exposure misclassification---self-reported DES exposure was substantially mismeasured in a validation study. But the study further revealed a complex interplay between the two phenomena. First, cluster size depended on the exposure that was subject to misclassification. Second, the degree of misclassification varied by cluster size, with large families experiencing the most severe misclassification. In this paper, we aim to understand the potential for bias in this setting and propose methods to correct for it. 

	To the best of our knowledge, misclassification has not been investigated in the context of potentially informative cluster size (ICS), or when misclassification is itself differential with respect to cluster size. In investigating this setting, we make a number of methodological contributions. First, we show that misclassification in an exposure related to cluster size can induce informativeness when cluster size would otherwise be non-informative. Second, we show that misclassification that depends on informative cluster size induces differential misclassification (with respect to outcomes)---and this can not only attenuate estimates of exposure associations but even inflate or reverse the sign of estimates. 
	Drawing on the literature for marginal inference under ICS,  
	we propose two frameworks to correct for misclassification: (1) an observed likelihood approach when one would have fit a joint marginalized model for cluster size and outcomes in the absence of misclassification, and (2) an expected estimating equations approach when one would have solved semi-parametric estimating equations. Although our focus is on marginal inference, a corollary of the observed likelihood approach is that it corrects also for misclassification in conditional models. 

	Section \ref{s:ICS3} reviews estimation under ICS, and Section \ref{s:misclass} describes the misclassification problem. We propose an observed likelihood framework in Section \ref{s:methods_lik} and an expected estimating equations framework in Section \ref{s:methods_eee}. In Sections \ref{s:sim3} and \ref{s:data3}, we evaluate the proposed methods in a simulation study and by applying them to motivating data from the Nurses Health Study II (NHSII) on DES exposure and ADHD among $N$$=$106,198 children to $K$$=$47,540 nurses. 

	\section{Estimation \& Informative Cluster Size in the Absence of Misclassification}
	\label{s:ICS3}
	Suppose we are interested in fitting a marginal model in a correlated data setting with $k=1,\ldots,K$ clusters, each of size $N_k$. Let $Y_{ki}$ be the outcome for the $i^{th}$ unit of the $k^{th}$ cluster ($i=1,\ldots,N_k$). Our primary goal is to estimate the effect of a cluster-level exposure $X_{1k}$. In the motivating study, this is DES exposure, which is binary and scalar. Let $X_{ki}=(X_{1k},X_{2ki})$, where $X_{2ki}$ is a vector that includes cluster- and unit-level covariates related to the outcome. Further let $Z_{k}=(X_{1k},Z_{2k})$, where $Z_{2k}$ is a vector of cluster-level covariates related to cluster size, which may share cluster-level covariates with $X_{2ki}$. 
	
	Typically, we take $N_k$ to be fixed. Specifying a mean model $g(\mu_{ki})=x\beta$, where $g(\cdot)$ is a link function and $\mu_{ki}=E[Y_{ki}|X_{ki}=x]$, estimation can proceed by assuming a marginally-specified GLMM \citep{heagerty2000marginalized} or by solving GEE \citep{liang1986longitudinal}.  
	
	Cluster size is often not fixed and may in fact be related to outcomes. Cluster size is \emph{informative} if $E[Y_{ki}|X_{ki}=x,N_k]\neq E[Y_{ki}|X_{ki}=x]$ \citep{hoffman2001within,williamson2003marginal}. 
	Under informative cluster size (ICS), joint marginalized models and weighted estimating equations have been proposed to extend standard approaches.

	\subsection{Joint Models of Outcome and Cluster Size}
	One approach to marginal inference is to specify a joint marginalized model (JMM) of outcome and cluster size \citep{mcgee2019informatively}, e.g.
	\begin{align}
	h\left(E[N_{k}|Z_{k}]\right)&=Z_k \alpha,  &g\left(E[Y_{ki}|X_{ki}]\right)&=X_{ki}\beta, \label{eqn:JMM1} \\
	h\left(E[N_{k}|Z_{k},b_k]\right)&=\Omega_k +\gamma \tilde{Z}_k b_k, 
	&g\left(E[Y_{ki}|X_{ki},b_k]\right)&=\Delta_{ki}+ \tilde{X}_{ki} b_k,  \label{eqn:JMM2} %
	\end{align}
	where $b_k\sim MVN(0,\Sigma)$, $h(\cdot)$ and $g(\cdot)$ are link functions, $\tilde{X}_{ki}$ is a vector subset of $X_{ki}$ and $\tilde{Z}_k$ a diagonal matrix whose elements are a subset of $Z_k$. In particular, the exposure of interest, $X_{1k}$, appears in both the model for the outcome and the model for cluster size.
	
	The models in (\ref{eqn:JMM1}) parameterize exposure associations marginally, yielding population-averaged interpretations, while the shared random vector $b_k$ specifies within-cluster dependence structure in (\ref{eqn:JMM2}) as in \cite{heagerty2000marginalized}. Moreover, $b_k$ permits dependence between outcomes and cluster size, and $\gamma$ controls the level of informativeness. Cluster size is informative when $\gamma\neq 0$.

	Given the marginal parameters $(\alpha,\beta)$, the implicitly defined parameters $\Omega_k$ and $\Delta_{ki}$ can be computed by solving integral equations (by iterated expectation; see supplementary material for details).
	We then make distributional assumptions and maximize the joint likelihood:
	\begin{align}
	\mathcal{L}(Y,N|X)&=
	\prod_{k=1}^{K} \int \mathcal{L}(N_k|b_k) \mathcal{L}(Y_k|b_k) dF(b_k), \label{eqn:condLike}
	\end{align}
	where we integrate over the random effects distribution $F(\cdot)$. 
	
	The marginalized framework ensures population-averaged parameter interpretations, but if interest is instead in cluster-specific parameters, setting $\Omega_{k}=Z_k \alpha$ and $\Delta_{ki}=X_{ki}\beta$ reduces to the typical conditional joint model 
	\citep{dunson2003bayesian,gueorguieva2005comments}.

	\subsection{Weighted Estimating Equations}

	Two weighting schemes have been proposed to extend GEE under ICS.
	The first solves the inverse cluster size weighted estimating equations (WEE):
	\begin{align}
	0&=\sum_{k=1}^K \frac{1}{n_k} \sum_{i=1}^{n_k}\frac{\partial \mu_{ki}}{\partial \beta}v_{ki}^{-1}(y_{ki}-\mu_{ki}), \label{eqn:WEE}
	\end{align}
	where $v_{ki}$ is the variance for the $i^{th}$ unit of the $k^{th}$ cluster. This is simply GEE with working independence and weights equal to $1/n_k$. The solution is consistent for $\beta^W$ defined by
	\begin{align}
	g\left(E[Y_{kI_k}|X_{kI_k}=x]\right)&=x\beta^W, \label{eqn:WEE_mod}
	\end{align}
	where $I_k$ is random index, with discrete uniform distribution on $\{1,\ldots,N_k\}$  \citep{williamson2003marginal,benhin2005mean}. 
	
	Alternatively, the independence estimating equations (IEE) approach solves the same equations as (\ref{eqn:WEE}) but with weights 1 instead of ${1}/{n_k}$. The solution is consistent for $\beta^{I}$ in
	\begin{align}
	g\left(\frac{E[N_k Y_{kI_k}|X_{kI_k}=x]}{E[N_k|X_{kI_k}=x]}\right)&=x\beta^{I}, \label{eqn:IEE_mod}
	\end{align}
	\citep{seaman2014review}.

	When cluster size is non-informative, $\beta^w=\beta^I$; under ICS, they differ \citep{seaman2014review}. We focus here on the WEE estimator, but the development pertains also to the IEE estimator. Details  can be found in the supplementary materials.

	\section{Exposure Misclassification When Cluster Size Depends on Exposure}
	\label{s:misclass}
	Consider a joint model of the form (\ref{eqn:JMM1})--(\ref{eqn:JMM2}) as a data generating mechanism, and  recall $X_{ki}=(X_{1k},X_{2ki})$ and $Z_{k}=(X_{1k},Z_{2k})$. 
	In practice we might not always observe the true exposure of interest, $X_{1k}$, and instead observe a misclassified version, $W_{1k}$. Because the exposure affects both outcomes and cluster size, na\"ive analyses---ones that replace $X_{1k}$ with $W_{1k}$---can misspecify both the outcome and size models. 
	
	Suppose $X_{1k}$ is binary and scalar, as is DES exposure. We consider two misclassification models---simple misclassification that depends only on the true exposure, e.g.:
	\begin{align}
	\text{logit}(E[W_{1k}|X_{1k}])&=\nu_0+\nu_1 X_{1k}, \label{eqn:misI}
	\end{align}
	and misclassification that depends on cluster size, e.g.:
	\begin{align}
	\text{logit}(E[W_{1k}|X_{1k},N_k])&=\nu_0'+\nu_1' X_{1k} +\nu_{2}'N_k+\nu_{3}'X_{1k}N_k. \label{eqn:misII}
	\end{align}
	The latter permits sensitivity and specificity (with respect to the true exposure) to be related to cluster size. For example, women who experienced fertility issues may have had more reason to learn about their mothers' pregnancies, hence exposures may be better measured in smaller clusters. This is motivated by the NHSII application, where sensitivity was substantially higher in smaller families (see Section \ref{s:data3}). 
	
	Finally, we assume $X_{1k}$ is available on a validation sample of clusters. Let $R_{k}=1$ indicate membership in the validation sample and $R_{k}=0$ indicate membership in the main sample. We observe $X_{1k}$ only when $R_{1k}=1$ but observe $W_{1k}$ for the entire sample.
	
	\subsection{Induced Informativeness}
	\label{ss:induced_ICS}
	Suppose cluster size is non-informative ($\gamma$=0 in expression \ref{eqn:JMM2}). In the absence of misclassification, we might fit the marginal outcome model in (\ref{eqn:JMM1}) via GEE. If we instead fit the na\"ive outcome model (replacing $X_{1k}$ with $W_{1k}$), not only is it subject to misclassification bias, but it is further subject to ICS whenever both cluster size and outcomes depend on an exposure that is mismeasured. Intuitively, cluster size and outcome are related through the true exposure, so the na\"ive model retains residual correlation not accounted for by the misclassified exposure (see supplementary material for theoretical details). We report evidence of induced informativeness via simulations in Section \ref{ss:sim_induced3}.

	\subsection{Induced Differential Misclassification}\label{ss:induced_differential}
	Suppose cluster size is informative ($\gamma \neq 0$) and that misclassification depends on size as in (\ref{eqn:misII}). Misclassification depends on a variable ($N_k$) that is itself related to the outcome: $E[Y_{kI_k}|X_{1k},X_{2kI_k},W_{1k}]\neq E[Y_{kI_k}|X_k,X_{2kI_k}]$,
	since $Y_{kI_k}$$\not\perp$$N_k|X_{1k},X_{2kI_k}$ and $W_{1k}$ depends on $N_k$. Intuitively, there is information about outcomes in $W_{1k}$ beyond that explained by $X_{1k}$. Thus when misclassification depends on cluster size that is itself informative, misclassification can be considered differential.

	\section{Proposed Method 1: Observed Likelihood}
	\label{s:methods_lik}
	If, in the absence of misclassification, one would have fit a joint model for cluster size and outcome---specified either marginally (JMM) for population-averaged interpretations, or conditionally for cluster-specific interpretations---then inference follows from a full likelihood specification. Under misclassification, the observed likelihood for a given cluster is:
	\begin{align}
	P(Y_k,N_k,W_{1k}) 
	&=\int P(Y_k,N_k|W_{1k},X_{1k})P(W_{1k}|X_{1k})P(X_{1k})dX_{1k}, \label{eqn:obslik_naive} 
	\end{align}
	where we omit notation for dependence on correctly observed covariates $\{X_{2k},Z_{2k}\}$ for simplicity. We assume $P(W_{1k}|Y_k,N_k,X_{1k})=P(W_{1k}|N_k,X_{1k})$, which allows misclassification to depend on cluster size:
	\begin{align}
	P(Y_k,N_k,W_{1k})&=\int P(W_{1k}|N_k,X_{1k}) P(Y_k,N_k|X_{1k})P(X_{1k})dX_{1k}. \label{eqn:obslik_N} 
	\end{align}

	We already specify $P(Y_k,N_k|X_{1k})$ in the absence of misclassification, so we need only  specify models for $P(W_k|N_k,X_{1k})$ and $P(X_{1k})$. Since $X_{1k}$ and $W_{1k}$ are binary in our setting, we can simply adopt logistic regression models and the integral reduces to a sum of two components. As above, $P(Y_k,N_k|X_{1k})$ is an integrated likelihood, although it is now a function of the $X_{1k}$ which is being marginalized over. See supplementary material for details.
	
	Given a validation sample, the observed likelihood is 
	\begin{align*}
	\prod_{k=1}^K \left[P(Y_k,N_k,W_{1k})\right]^{(1-R_k)}\left[P(Y_k,N_k,W_{1k},X_{1k})\right]^{R_k}
	\end{align*}
	\citep{spiegelman2000estimation}. Inference follows by taking the inverse of the information.
	
	\subsection{Exposure Misclassification and Non-ICS}
	\label{ss:GLMM3}
	We can similarly adopt an observed-likelihood approach to inference in the special case when one would have simply fit a (conditionally- or marginally-specified) GLMM---that is, when we assume no ICS ($\gamma$=0). However we cannot simply replace $P(Y_k,N_k|X_{1k})$ with $P(Y_k|X_{1k})$, due to the relationship between $\{W_{1k},X_{1k}\}$ and $N_k$. Instead we treat size as fixed and write:
	\begin{align}
	P(Y_k,W_{1k}|N_k) 
	&=\int P(Y_k|X_{1k})P(W_{1k}|X_{1k},N_k)P(X_{1k}|N_k)dX_{1k}, \label{eqn:obslik_naive_glmm2} 
	\end{align}
	by non-differential misclassification, $P(Y_k|W_{1k},X_{1k},N_k)=P(Y_k|X_{1k},N_k)$, and non-ICS, $P(Y_k|X_{1k},N_k)=P(Y_k|X_{1k})$. So analysts need to adjust for cluster size in the exposure model $P(X_{1k}|N_k)$, even if misclassification does not depend on size.

	\section{Proposed Method 2: Expected Estimating Equations}
	\label{s:methods_eee}

	Estimation in the absence of misclassification can proceed by solving 0 = $\sum_{k=1}^K U_{k}(\beta;Y_{k},X_{1k},X_{2k})$ for some unbiased complete-data estimating function $U_{k}(\beta;Y_{k},X_{1k},X_{2k})$, as in GEE. However, if one observes $W_{1k}$ in place of $X_{1k}$, one cannot compute $U_{k}(\beta;Y_k,X_{1k},X_{2k})$. Treating $N_k$ as fixed, \cite{wang2000expected} and \cite{wang2008expected} instead take as estimating function the expectation of the complete-data estimating function conditional on ($Y_{k},W_{1k},X_{2k}$). That is, one solves expected estimating equations (EEE) $0=\sum_{k=1}^K E_{X_{1k}}[U_{k}(\beta;Y_k,X_{1k},X_{2k})|Y_{k},W_{1k},X_{2k}]$, which are unbiased by iterated expectation and depend only on observed data.

	In our context, $N_k$ is no longer fixed. As such we propose to augment the expected estimating equations approach to incorporate cluster size. 
	%
	%
	%
	%
	%
	Under ICS, we take WEE (\ref{eqn:WEE}) as complete-data estimating equations. In Supplementary Appendix B, we show
	\begin{align}
	&E_{Y_k,W_{1k},X_{2ki},N_k}\left[\frac{1}{N_k}\sum_{i=1}^{N_k}E_{X_{1k}}\left[\frac{\partial \mu_{ki}}{\partial \beta}v_{ki}^{-1}(Y_{ki}-\mu_{ki})|Y_{ki},W_{1k},X_{2ki},N_k\right]\right]=0, \nonumber
	\end{align}
	so that unbiased estimating equations can be constructed as
	\begin{align}
	&\sum_{k=1}^K \frac{1}{n_k} \sum_{i=1}^{n_k}E_{X_{1k}}\left[\frac{\partial \mu_{ki}}{\partial \beta}v_{ki}^{-1}(y_{ki}-\mu_{ki})|y_{ki},w_{ki},x_{2ki},n_k\right]=0, \label{eqn:eee_WEE}
	\end{align}
	which depend on observed outcome, covariates and cluster size. Proving unbiasedness (see Supplementary Appendix B) requires marginalizing over random cluster size and hence the derivation appeals to random indices as in (\ref{eqn:WEE_mod}), but the result is a straightforward extension of the approach of \cite{wang2000expected} and \cite{wang2008expected}.
	
	%
	
	\subsection{Practical Considerations}
	Evaluating the expectations in expression (\ref{eqn:eee_WEE}) involves computing:
	\begin{align}
	P(X_{1k}|Y_{ki},W_{1k},N_k)&=\frac{P(W_{1k}|Y_{ki},X_{1k},N_k)P(Y_{ki},N_k|X_{1k})P(X_{1k})}{\int P(W_{1k}|X_{1k},N_k)P(Y_{ki},N_k|X_{1k})P(X_{1k})dF(X_{1k})}, \label{eqn:decomp}
	\end{align}
	again omitting notation for dependence on $X_{2ki}$ for simplicity. Even after assuming non-differential misclassification, $P(W_{1k}|Y_{ki},X_{1k},N_k)=P(W_{1k}|X_{1k},N_k)$, this requires specifying the joint distribution of outcome and cluster size, $P(Y_{ki},N_k|X_{1k})$. If one were willing to make the necessary assumptions, one could fully specify distribution (\ref{eqn:decomp}) and guarantee consistent estimation \citep{wang2008expected}.
	

	Specifying $P(Y_{ki},N_k|X_{1k})$ poses a significant challenge in practice. In the absence of misclassification, a key advantage of a WEE analysis over the JMM approach is that one need not model the outcome-size relationship. If one were unwilling to make distributional assumptions in the absence of misclassification, one might be reluctant to make those same assumptions to address misclassification. In light of this, we propose a pragmatic path forward.

	As a practical alternative to full likelihood specification, one might instead specify a model for $P(X_{1k}|Y_{ki},W_{1k},N_k)$ directly, without regard to the decomposition given by (\ref{eqn:decomp}). This raises the possibility of model incompatibility---that is, the assumed form may not respect the model structure in (\ref{eqn:decomp}). As such, one is no longer guaranteed consistency---but the specification may nevertheless be a reasonable approximation in practice. Since $X_{1k}$ is binary, one option would be to fit a generalized linear model which incorporates $\{Y_{ki},W_{1k},N_k\}$ with a reasonable level of flexibility. For example, let $\mu^X_{ki}$ and $v^X_{ki}$ be mean and variance of $X_{1k}$ given $\{Y_{ki},W_{1k},N_k\}$, and specify $g(\mu^X_{ki})=Z^X_{ki}\xi$ where $Z^{X}_{ki}$ is a covariate vector incorporating $\{Y_{ki},W_{1k},N_k\}$.

	In the main sample, the approximate expected estimating equations are:
	\begin{align*}
	0&=\sum_{k=1}^K \frac{1}{n_k} \sum_{i=1}^{n_k}E_{X_{1k}}\left[\frac{\partial \mu_{ki}}{\partial \beta}v_{ki}^{-1}(y_{ki}-\mu_{ki})|y_{ki},w_{1k},n_k\right] \\
	0&=\sum_{k=1}^K \sum_{i=1}^{n_k}E_{X_{1k}}\left[\frac{\partial \mu^X_{ki}}{\partial \xi}(v_{ki}^X)^{-1}(X_{1k}-\mu^X_{ki})|y_{ki},w_{1k},n_k\right],
	\end{align*}
	and we drop the expectation in the validation sample. If we adopt a logistic link for the outcome and the exposure models, we solve:
	\begin{align}
	0&=\sum_{k=1}^K \frac{1}{n_k} \sum_{i=1}^{n_k}\left\lbrace R_{k}\left[x_{ki}^T(y_{ki}-\mu_{ki}) \right]\right.\nonumber\\
	&~~~~~~~~~~~~~~\left. +(1-R_{k})\left[(x_{ki}^{1})^T(y_{ki}-\mu_{ki}^1)\mu^X_{ki}+(x_{ki}^{0})^T(y_{ki}-\mu_{ki}^0)(1-\mu^X_{ki}) \right] \right\rbrace \label{eqn:AEE1b}\\
	0&=\sum_{k=1}^K \sum_{i=1}^{n_k} R_{k}\left[(z^{X}_{ki})^T(x_{1k}-\mu^X_{ki}) \right], \label{eqn:AEE2b}
	\end{align}
	where $\{x_{ki}^1,x_{ki}^0\}$ are covariate vectors $x_{ki}$ but with exposure $X_{1k}$ set to $\{1,0\}$ respectively; $\{\mu_{ki}^1,\mu_{ki}^0\}$ are outcome means as $\mu_{ki}$ but based on $X_{1k}=\{0,1\}$; and $R_k$ is an indicator of inclusion in the validation sample. 
	
	The main sample contribution ($R_k$=0) in expression (\ref{eqn:AEE2b}) reduces to zero, so estimation can proceed via a plug-in estimator: first fit the exposure model to the validation sample, then substitute the fitted values in the estimating equations for the outcome model. As such, one might entertain the full range of statistical and machine learning tools in the first stage exposure model, but this is beyond the scope of this paper. We compute standard errors via non-parametric bootstrap, resampling $\sum_{k=1}^K R_{k}$ clusters with replacement from the validation sample and $K-\sum_{k=1}^K R_{k}$ from the main sample. 

	
	
	\section{Simulation}
	\label{s:sim3}
	We conducted a series of simulations to: (1) investigate the interplay between ICS and misclassification with respect to bias, and (2) evaluate the extent to which the proposed methods correct for such bias and provide valid inference.
	
	
	\subsection{Generating Misclassified Data}
	\label{ss:sim_gendat3}
	We simulated $R$$=$2,000 datasets from a JMM with shared random intercepts (whose variance depends on exposure).
	For clusters $k=1,\dots,8000$:
	\begin{enumerate}
		\item Draw cluster-level covariate $X_{1k}\sim \text{Bernoulli}(0\text{.}25)$ and let $Z_k=(1,X_{1k})$. 
		\item Draw random effects  $b_k\sim \text{N}(\sigma_0^2)$ if $X_{1k}=0$ or $b_k\sim \text{N}(\sigma_1^2)$ if $X_{1k}=1$. 
		\item Set $\lambda_k=\text{exp}(\Omega_k +\gamma_0 b_{0k} (1-X_{1k}) +\gamma_1 b_{1k}X_{1k}) $ where $\Omega_k=Z_{k}\alpha$. 
		\item Draw cluster size $N_k|Z_k,b_k \sim \text{Poisson}(\lambda_k)+1$.
		\item For $i=1$ to $i=N_k$: 
		\begin{enumerate}
			\item Draw unit-level covariate $X_{2ki}\sim \text{Normal}(0,1)$, let $X_{ki}=(1,X_{1k},X_{2ki})$.
			\item Set $\mu_{ki}$$=$$\text{expit}(\Delta_{ki} + b_{0k} (1-X_{1k}) + b_{1k}X_{1k} )$, solving  $\text{expit}(X_{ki}\beta)$$=$$\int \mu_{ki} dF(b_k)$ for $\Delta_{ki}$. 
			\item Draw outcomes $Y_{ki}|X_{ki},b_k \sim \text{Bernoulli}(\mu_{ki})$.
		\end{enumerate}
	\end{enumerate}
	
	We set $\alpha$$=$$(0.6,-0.2)^T$, $\beta$$=$$(-4.0,0.5,0.2)^T$, $\sigma_0$=2.0, $\sigma_1$=1.5, and $\gamma_0$=$\gamma_1$=$\gamma$. We considered non-ICS ($\gamma$=0), as well as fairly strong informativeness in two directions: $\gamma$=$-$0.25 and $\gamma$=0.25. The mean cluster size was 2.74, and 74--75\% of clusters had no more than three members. For $\gamma$=$-$0.25, outcome prevalence was 4.3\% in single-member clusters and 1.3\% in three-member clusters; for $\gamma$=0.25, it was 0.7\% in single-member clusters and 1.8\% in three-member clusters. See supplementary material for full distributions.

	We generated misclassified exposure $W_{1k}$ first under (i) simple misclassification, where sensitivity and specificity with respect to $X_{1k}$ were fixed at 75\% and 85\%. Then, motivated by the NHSII, we considered misclassification models that depended on cluster size, where sensitivity and specificity (ii) \textit{both decreased} with cluster size, (iii) \textit{both increased} with cluster size, and (iv) mimicked the \textit{observed misclassification pattern} in the NHSII where sensitivity decreased with family size and specificity remained high. See Table \ref{tab:sim_mis} for details. 

	
	\begin{table}[htbp!]
		\centering
		\caption{Simulated misclassification: sensitivity and specificity (in \%) with respect to true exposure in four misclassification models. Sensitivity and specificity (i) remain fixed with respect to cluster size $N_k$, (ii) decrease in $N_k$, (iii) increase in $N_k$ or (iv)  reflect the observed misclassification patter. \label{tab:sim_mis}}
		\begin{tabular}{lccccccccc}
			\toprule
			&\multicolumn{2}{c}{Simple} & \multicolumn{6}{c}{Size-Dependent} \\
			\cmidrule(l{10pt}){2-3} \cmidrule(l{10pt}){4-9}
			& \multicolumn{2}{c}{(i) Fixed} & \multicolumn{2}{c}{(ii) Decrease} &  \multicolumn{2}{c}{(iii) Increase} &  \multicolumn{2}{c}{(iv) Observed}\\
			\cmidrule(l{10pt}){2-3}\cmidrule(l{10pt}){4-5}\cmidrule(l{10pt}){6-7} \cmidrule(l{10pt}){8-9}
			$N_k$~~~~~& Sens & Spec & Sens & Spec& Sens & Spec& Sens & Spec \\
			\midrule
			1 & 75 & 85 & 95 & 95 & 54 & 54 & 71 & 99 \\ 
			2 & 75 & 85 & 90 & 90 & 70 & 70 & 65 & 99 \\ 
			3 & 75 & 85 & 82 & 82 & 82 & 82 & 58 & 99 \\ 
			4 & 75 & 85 & 69 & 69 & 90 & 90 & 50 & 99 \\ 
			\bottomrule 
		\end{tabular}
	\end{table}

	We consider a hypothetical study where the true exposure, $X_{1k}$, is observed only in a validation subsample of $1,600$ clusters (20\%). 
	
	\subsection{Analyses}
	\label{ss:sim_analyses3}

	In each simulated dataset, we compared the performance of the JMM and WEE estimators based on the true model (based on true exposure $X_{1k}$), the na\"ive model (based on the misclassified $W_{1k}$), the validation estimators (based on true exposure in validation set only), as well as the proposed corrections. For JMM, we applied the observed likelihood correction, where we adopted the correct form of the misclassification model. For WEE, we applied four EEE corrections of increasing complexity. These model the conditional distribution of $X_{1k}$ via: main effects of $Y_{ki},W_{1k},X_{2ki}$ (EEE1); main effects of $Y_{ki},W_{1k},X_{2ki}$ as well as $N_k$ (EEE2); an interaction and main effects for $Y_{ki}$ and $W_{1k}$, and a main effect for $X_{2ki}$ (EEE3); interactions and main effects for $Y_{ki},W_{1k},N_{k}$ plus main effect for $X_{2ki}$ (EEE4). Wald-type 95\% confidence intervals were constructed for EEE estimators via bootstrap standard errors based on 50 resamples.  (See supplementary material for IEE analyses.)

	
	\subsection{Results}
	\label{ss:sim_results3}
	
	Table \ref{tab:simB1} summarizes operating characteristics for estimates of the exposure effect, $\beta_1$. We report results for $(\beta_0,\beta_2)$ in supplementary  Tables.
	
	\begin{table}
		\centering
		
		\caption{\label{tab:simB1}Simulation results: mean, standard deviation (SD) and 95\% confidence interval coverage (Cvg) for estimates of exposure effect ($\beta_1=0.5$) across $R$=2,000 simulated datasets in twelve informativeness/misclassification pairings, described in Section \ref{ss:sim_gendat3}. Each dataset contains $K$=8,000 clusters, with 1,600 clusters drawn as a validation sample. True is based on $X_{1k}$ in full data; Mis is based on $W_{1k}$ in full data; Valid is based on $X_{1k}$ in the validation sample only.} 
		\scalebox{0.8}{
			\begin{tabular}[htbp!]{ccrrrrrrrrrrrr}
				\toprule
				ICS& Size-Dep. &   & \multicolumn{4}{c}{JMM} & \multicolumn{7}{c}{WEE}\\
				\cmidrule{4-7} \cmidrule(l{8pt}){8-14} 
				($\gamma$) & Misclass. &  & True & Mis & Valid & ObsLik & True & Mis & Valid & EEE1 & EEE2 & EEE3 & EEE4 \\
				\midrule
				No & No (i)& Mean  & 0.50 & 0.28 & 0.50 & 0.50 & 0.50 & 0.28 & 0.49 & 0.50 & 0.50 & 0.50 & 0.50\\
				(0)&(Simple)& SD & 0.12 & 0.12 & 0.27 & 0.18 & 0.12 & 0.12 & 0.28 & 0.22 & 0.22 & 0.22 & 0.24\\
				& & Cvg & 95 & 56 & 95 & 94 & 95 & 57 & 95 & 95 & 94 & 94 & 94\\
				\cmidrule{1-14}
				No& Yes (ii)& Mean & 0.50 & 0.34 & 0.50 & 0.50 & 0.50 & 0.28 & 0.49 & 0.53 & 0.52 & 0.53 & 0.50\\
				(0)&(Decrease) & SD & 0.12 & 0.12 & 0.27 & 0.16 & 0.12 & 0.11 & 0.28 & 0.24 & 0.22 & 0.24 & 0.21\\
				& & Cvg & 95 & 70 & 95 & 95 & 95 & 54 & 95 & 94 & 95 & 94 & 94\\
				\cmidrule{1-14}
				No&Yes (iii)& Mean & 0.50 & 0.24 & 0.50 & 0.50 & 0.50 & 0.23 & 0.49 & 0.47 & 0.47 & 0.47 & 0.50\\
				(0)& (Increase)& SD & 0.12 & 0.12 & 0.27 & 0.20 & 0.12 & 0.12 & 0.28 & 0.22 & 0.22 & 0.22 & 0.26\\
				& & Cvg & 95 & 38 & 95 & 94 & 95 & 37 & 95 & 94 & 94 & 94 & 94\\
				\cmidrule{1-14}
				No&Yes (iv)& Mean & 0.50 & 0.41 & 0.50 & 0.50 & 0.50 & 0.41 & 0.49 & 0.50 & 0.50 & 0.50 & 0.50\\
				(0)& (Observed)& SD & 0.12 & 0.14 & 0.27 & 0.16 & 0.12 & 0.14 & 0.28 & 0.20 & 0.20 & 0.20 & 0.21\\
				& & Cvg & 95 & 92 & 95 & 95 & 95 & 91 & 95 & 95 & 94 & 95 & 94\\
				\cmidrule{1-14}
				Yes&No (i)& Mean & 0.50 & 0.29 & 0.49 & 0.49 & 0.50 & 0.28 & 0.49 & 0.60 & 0.57 & 0.60 & 0.49\\
				(--0.25)& (Simple)& SD & 0.13 & 0.13 & 0.30 & 0.19 & 0.14 & 0.14 & 0.32 & 0.24 & 0.24 & 0.25 & 0.27\\
				& & Cvg & 95 & 63 & 95 & 95 & 95 & 65 & 95 & 92 & 94 & 92 & 95\\
				\cmidrule{1-14}
				Yes&Yes (ii)& Mean & 0.50 & 0.12 & 0.49 & 0.50 & 0.50 & 0.07 & 0.49 & 0.70 & 0.60 & 0.68 & 0.50\\
				(--0.25)&(Decrease)& SD & 0.13 & 0.13 & 0.30 & 0.16 & 0.14 & 0.13 & 0.32 & 0.24 & 0.23 & 0.23 & 0.21\\
				& & Cvg & 95 & 15 & 95 & 95 & 95 & 9 & 95 & 84 & 92 & 86 & 96\\
				\cmidrule{1-14}
				Yes&Yes (iii)& Mean & 0.50 & 0.43 & 0.49 & 0.49 & 0.50 & 0.41 & 0.49 & 0.55 & 0.55 & 0.54 & 0.49\\
				(--0.25)&(Increase)& SD & 0.13 & 0.12 & 0.30 & 0.23 & 0.14 & 0.13 & 0.32 & 0.27 & 0.27 & 0.27 & 0.31\\
				& & Cvg & 95 & 92 & 95 & 95 & 95 & 90 & 95 & 94 & 94 & 94 & 96\\
				\cmidrule{1-14}
				Yes&Yes (iv)& Mean & 0.50 & 0.48 & 0.49 & 0.50 & 0.50 & 0.49 & 0.49 & 0.57 & 0.56 & 0.57 & 0.49\\
				(--0.25)& (Observed)& SD & 0.13 & 0.15 & 0.30 & 0.17 & 0.14 & 0.15 & 0.32 & 0.23 & 0.23 & 0.23 & 0.24\\
				& & Cvg & 95 & 95 & 95 & 95 & 95 & 96 & 95 & 94 & 94 & 94 & 95\\
				\cmidrule{1-14}
				Yes&No (i)& Mean & 0.50 & 0.27 & 0.51 & 0.50 & 0.50 & 0.28 & 0.50 & 0.37 & 0.40 & 0.38 & 0.51\\
				(0.25)&(Simple)& SD & 0.11 & 0.10 & 0.23 & 0.14 & 0.11 & 0.11 & 0.25 & 0.21 & 0.20 & 0.21 & 0.20\\
				& & Cvg & 94 & 36 & 95 & 95 & 94 & 45 & 95 & 91 & 92 & 92 & 95\\
				\cmidrule{1-14}
				Yes&Yes (ii)& Mean & 0.50 & 0.69 & 0.51 & 0.50 & 0.50 & 0.81 & 0.50 & 0.30 & 0.40 & 0.36 & 0.50\\
				(0.25)&(Decrease)& SD & 0.11 & 0.09 & 0.23 & 0.15 & 0.11 & 0.10 & 0.25 & 0.26 & 0.22 & 0.24 & 0.19\\
				& & Cvg & 94 & 49 & 95 & 95 & 94 & 14 & 95 & 88 & 92 & 92 & 95\\
				\cmidrule{1-14}
				Yes&Yes (iii)& Mean & 0.50 & 0.00 & 0.51 & 0.51 & 0.50 & -0.04 & 0.50 & 0.34 & 0.34 & 0.40 & 0.50\\
				(0.25)&(Increase)& SD & 0.11 & 0.10 & 0.23 & 0.13 & 0.11 & 0.11 & 0.25 & 0.17 & 0.17 & 0.16 & 0.19\\
				& & Cvg & 94 & 0 & 95 & 95 & 94 & 0 & 95 & 86 & 86 & 91 & 94\\
				\cmidrule{1-14}
				Yes&Yes (iv)& Mean & 0.50 & 0.28 & 0.51 & 0.50 & 0.50 & 0.27 & 0.50 & 0.39 & 0.40 & 0.39 & 0.51\\
				(0.25)&(Observed)& SD & 0.11 & 0.13 & 0.23 & 0.14 & 0.11 & 0.14 & 0.25 & 0.19 & 0.19 & 0.19 & 0.20\\
				& & Cvg  & 94 & 62 & 95 & 96 & 94 & 62 & 95 & 90 & 91 & 90 & 95\\		
				\bottomrule
		\end{tabular}}
	\end{table}
	
	Na\"ive JMM estimates of $\beta_1$---based on the misclassified exposure---were substantially biased across the board. Under simple misclassification, na\"ive estimates were attenuated by roughly 44\%, and this was fairly stable with respect to informativeness (42--46\%). By contrast, under size-dependent misclassification, bias was amplified by the presence of informativeness: when sensitivity/specificity decreased with cluster size, estimates were attenuated by 32\% under non-ICS ($\gamma$=0), but estimates were much more attenuated (76\%) when $\gamma$=$-$0.25, and were actually inflated by 38\% when $\gamma$=0.25. When sensitivity/specificity instead increased with cluster size, informativeness shifted bias in the opposite directions. Na\"ive WEE estimates followed the same pattern, but generally exhibited more bias---in particular, when concordance increased with cluster size and $\gamma$=0.25, we observed downward bias of more than 100\%---on average reversing the sign of the estimates relative to the truth.
	%
	%

	The observed likelihood estimators were approximately unbiased and highly efficient, yielding standard errors 24--48\% smaller than those of the validation-only estimators. 
	
	The most complex EEE specification (EEE4) produced approximately unbiased estimates across all scenarios. The simpler models (EEE1---EEE3) yielded little to no bias under non-ICS, but only partially corrected for bias elsewhere. Indeed the choice of EEE specification revealed a bias-variance tradeoff: although the most complex model yielded full bias correction, it did so at the cost of increased variability. In some cases (e.g., $\gamma$=$\{0,-0.25\}$ and simple misclassification), this increase in variability resulted in a lower MSE for the simpler models (see Table \ref{tab:supp_simB1}). Overall the results suggest that fairly flexible specification of the conditional distribution of the exposure is required to fully correct for bias.

	Table \ref{tab:simB1} also reports 95\% confidence interval coverage. All estimators that were approximately unbiased achieved near nominal coverage.

	%
	%
	%

	\subsection{Induced Informativeness}
	\label{ss:sim_induced3}
\begin{figure}[htbp]
		\centering 
		\subfloat[~Complete-data models]{\includegraphics[width=0.4\linewidth]{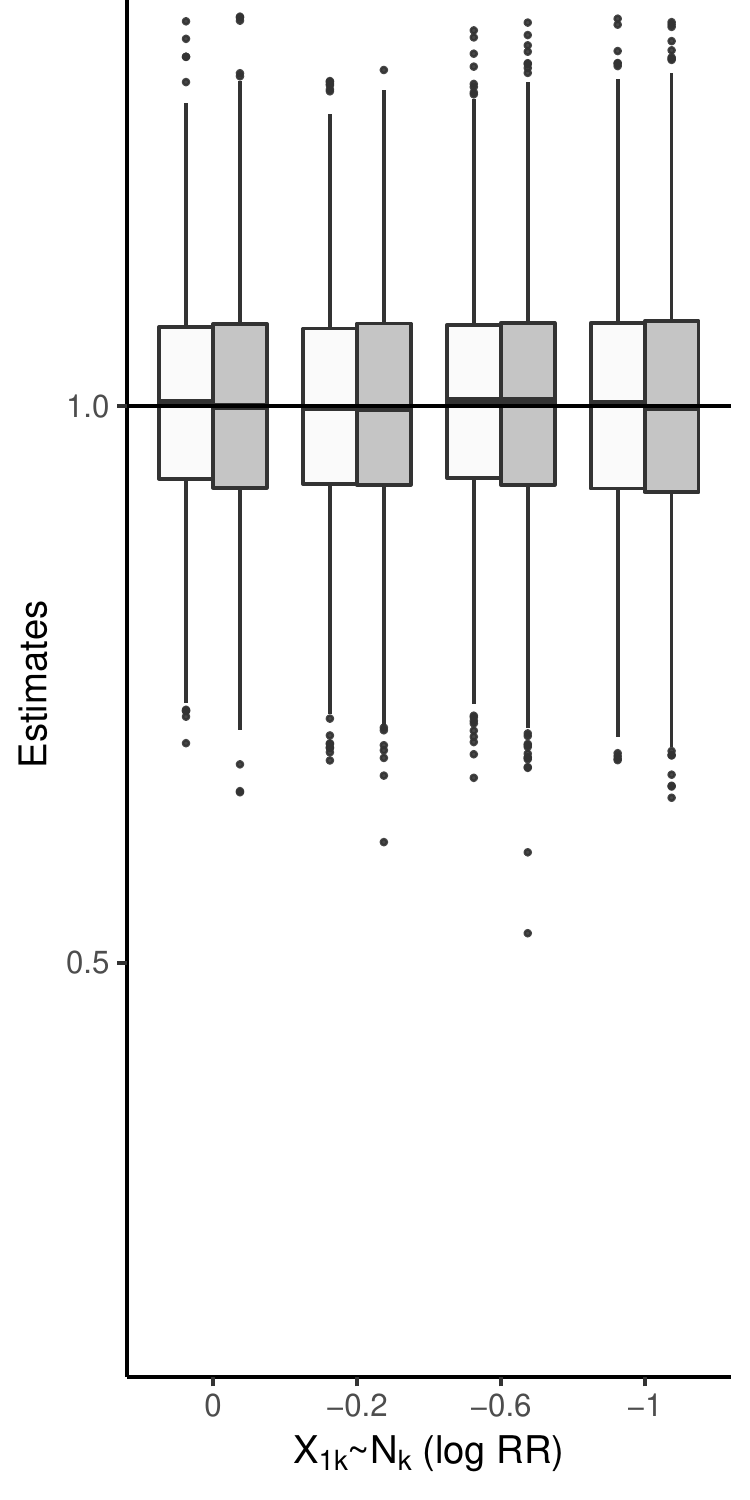}}
		\subfloat[~Naive models]{\includegraphics[width=0.4\linewidth]{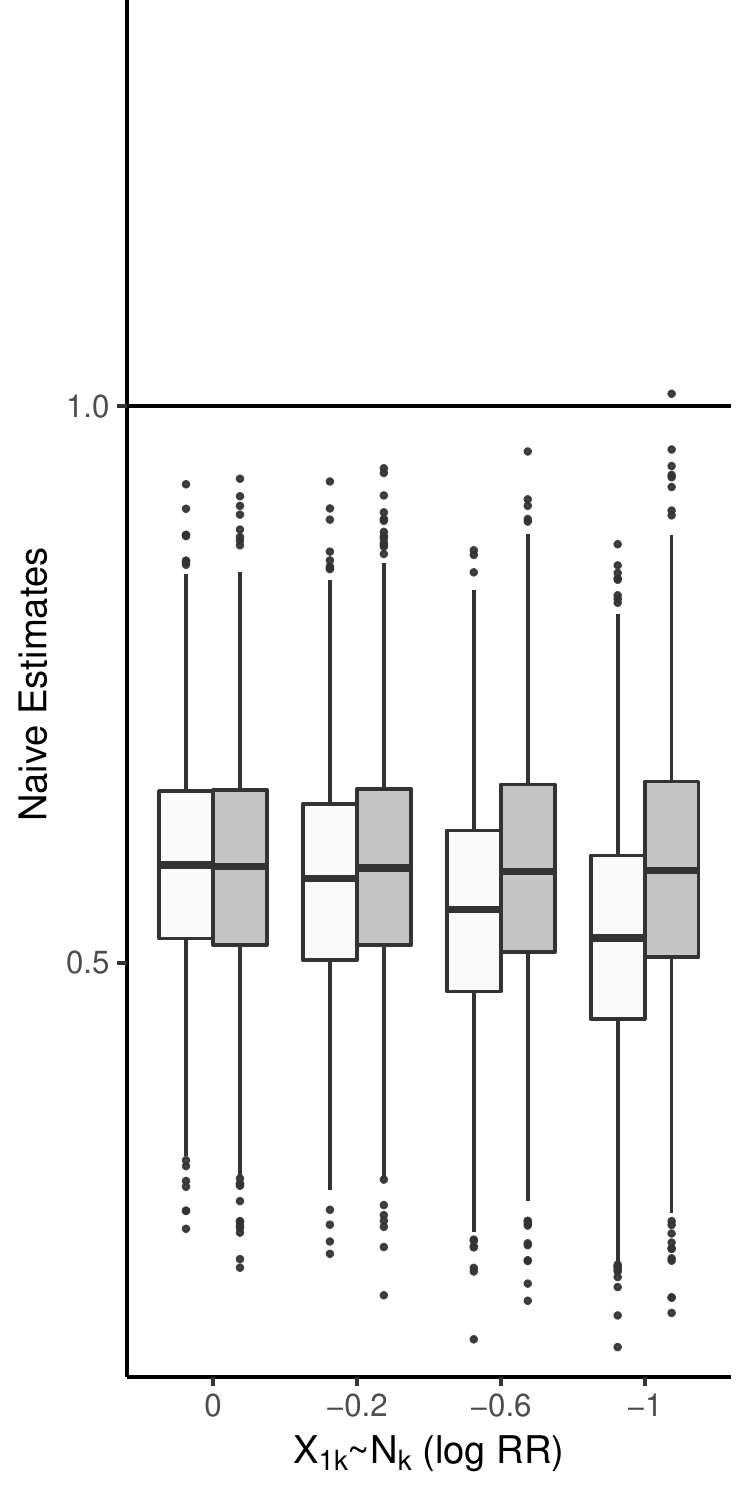}}

		\caption{Induced informativeness: IEE (white) and WEE (gray) estimates of misclassified exposure effect across $R$=2,000 simulated datasets for different exposure-size associations. Left panel is complete-data model estimates; right panel is na\"ive (misclassified) model estimates.  
			See supplementary material for data generation details.}
		\label{fig:mis_box}
	\end{figure}

	We ran a supplementary simulation to first investigate the effect of induced misclassification as outlined in Section \ref{ss:induced_ICS}. We generated data as in Section \ref{ss:sim_gendat3} with $\gamma$=0 so that the data-generating model exhibited no ICS, and we set $\beta_1$=1 and $\alpha_1$=$\{0.0,-0.2,-0.6,-1.0\}$. In Figure \ref{fig:mis_box}, we show that (even under simple misclassification as in \ref{eqn:misI}) the na\"ive IEE and WEE estimators diverge as the exposure-size association increases---all while the true model estimators (based on $X_{1k}$) remain stable. Since the IEE and WEE estimators are consistent for the same parameter under non-ICS, this reflects induced informativeness. 
	
	Second, we studied the performance of correction methods on non-ICS analyses. We set $\alpha_1$=$-$1 to reflect the strongest induced informativeness in Figure \ref{fig:mis_box}. Since the true model is not subject to ICS, one might fit a marginally-specified GLMM (for the outcome only), and we applied the observed likelihood approach of Section \ref{ss:GLMM3}. This is a standard correction approach, modified only to acknowledge dependence on cluster size in both the exposure and the misclassification models. Similarly, since the true model is not subject to ICS, one might simply fit IEE, which is equivalent to GEE with working independence. We then applied the proposed EEE framework. Since \cite{wang2000expected} adopted GEE with working independence, this simply represents an approximation to the standard EEE approach, again taking care to acknowledge dependence on cluster size where appropriate.  All correction methods produced full bias correction; see supplementary Tables for details.

	\section{Data Application: The Nurses Health Study II}
	\label{s:data3}
	\subsection{Study Population}
	\label{ss:data_pop3}
	The motivating study of the effect of in-utero DES exposure on next-generation ADHD diagnoses employed data from the Nurses Health Study II, which consisted of 47,540 female nurses, aged 25--42 in 1989 (see \citet{kioumourtzoglou2018association} for details).
	The main study units were the nurses' 106,198 children, clustered within families. We see in Table \ref{tab:ICS_tab} that the outcome of interest, ADHD diagnosis (as reported by the nurses), was most common in smaller families. ADHD prevalence ranged from 5.62\% among only-children to 4.14\% among those from the largest families, suggesting potentially informative cluster size.
	
	\begin{table}[htbp!]
		\centering
		\caption{Outcome and exposure prevalence by number of children (cluster size) in the Nurses Health Study II. $N_k$ refers to number of children (cluster size); DES refers to (misclassified) diethylstilbestrol exposure as reported by nurses. Sensitivity and specificity are with respect to mother-reported DES exposure. \label{tab:ICS_tab}}
		\begin{tabular}{lrccccc}
			\toprule
			& \multicolumn{3}{c}{Full sample} & \multicolumn{3}{c}{Validation sample} \\
			\cmidrule(l{8pt}){2-4} \cmidrule(l{8pt}){5-7}
			$N_k$ & Nurses & ADHD & DES & Nurses & Sensitivity & Specificity \\ 
			& \multicolumn{1}{c}{(No.)} & \multicolumn{1}{c}{(\%)} & \multicolumn{1}{c}{(\%)}  & \multicolumn{1}{c}{(No.)} & \multicolumn{1}{c}{(\%)} & \multicolumn{1}{c}{(\%)} \\
			\midrule
			1 & 8,791 & 5.62 & 2.37 & 3,416 & 69.8 & 99.6 \\ 
			2 & 23,608 & 5.44 & 1.88 & 9,383  & 62.2 & 99.7 \\ 
			3 & 11,444 & 5.44 & 1.39 & 4,491  & 59.8 & 99.9 \\ 
			4+ & 3,697 & 4.14 & 1.38 & 1,502  & 50.0 & 100.0 \\ 
			\bottomrule
		\end{tabular}
	\end{table}
	
	Nurses reported also whether they were exposed to DES in-utero. This nurse-reported exposure was rare and its prevalence decreased with family size: 2.37\% among nurses with one child to 1.38\% among those with four or more. Since DES was administered before the nurses' actual births,  nurses' reported exposure ($W_{1k}$) was naturally subject to misclassification. The study later collected a large validation sample of 18,792 (40\%) nurses, whose mothers reported their own DES use during pregnancy with the corresponding daughter (i.e., nurse). We take mother-reported exposure to be the truth ($X_{1k}$). 
	
	Table \ref{tab:ICS_tab} reveals that misclassification depends strongly on cluster size: while the specificity of nurse-reported DES exposure (with respect to mother-reported exposure) remained above 99\%, the sensitivity decreased from 70\% among nurses with one child down to 50\% among those with four or more. This is consistent with the hypothesis that nurses who experienced difficulty bearing children may be more aware of their mother's pregnancy history.

	\subsection{Analyses}
	\label{ss:data_analyses3}
	\cite{kioumourtzoglou2018association} estimated the marginal effect of DES on ADHD using the nurse-reported exposure. We first estimated this  effect by fitting logistic regression models via JMM and WEE that contained nurse-reported exposure, along with covariates: smoking status and year of birth. For the JMM, we adopted a Poisson model for cluster size, adjusting for the same variables and incorporating shared random intercepts whose variance depended on DES exposure. 
	
	We then applied the proposed corrections. For the JMM analysis, we adopted the observed likelihood approach of Section \ref{s:methods_lik}, where the misclassification model included cluster size, and the exposure model included the covariates. For the WEE analysis, we adopted the EEE approach of Section \ref{s:methods_eee}, incorporating a three-way interaction for outcome, size and nurse-reported exposure, as well as covariates, in the conditional exposure model. As a sensitivity analysis, we also considered a simpler EEE specification that excluded interactions in the conditional exposure model.  See Supplementary Section \ref{sec:model_spec} for full model specifications.


	
	\subsection{Results}
	\label{ss:data_results}
	We report marginal parameter estimates and confidence intervals (CI) in Table \ref{tab:app_res}. For the JMM analyses, the na\"ive and corrected estimates of the odds ratio for DES on ADHD were similar (1.42, 95\% CI [1.16,1.73]; 1.44, 95\% CI [1.18, 1.75]). Similarly, the na\"ive and corrected WEE estimates differed only slightly (1.39 [1.13, 1.71]; 1.33 [1.03, 1.71]). In simulations, we saw that overly simple EEE specifications  only partially corrected for bias---we saw this reflected here, where a simpler EEE specification (excluding interactions) yielded an estimate much closer to the na\"ive (1.38 [1.08, 1.75]). Na\"ive and corrected estimates of covariate odds ratios were also nearly identical. 

	\begin{table}[htbp!]
		\centering
		\caption{Marginal parameter estimates across analyses of ADHD diagnosis in $N$$=$106,198 children. Na\"ive analyses use nurse-reported (misclassified) exposure. WEE is inverse cluster size weighted estimating equations; JMM is joint marginalized model; Corrected WEE applies the EEE correction; Corrected JMM applies the observed likelihood correction; Est refers to estimates; CI refers to confidence intervals. \label{tab:app_res}}
		\scalebox{0.9}{
			\begin{tabular}{lcccccccc}
				\toprule
				&\multicolumn{4}{c}{Na\"ive} &\multicolumn{4}{c}{Corrected} \\
				\cmidrule(l{8pt}){2-5}\cmidrule(l{8pt}){6-9} 
				& \multicolumn{2}{c}{WEE} & \multicolumn{2}{c}{JMM}& \multicolumn{2}{c}{WEE} & \multicolumn{2}{c}{JMM}\\
				& Est &  95\% CI & Est &  95\% CI & Est &  95\% CI & Est &  95\% CI  \\
				\midrule
				\emph{Family Size}\hfill \\
				{Baseline Rate} &  &  & 1.19 & (1.18, 1.21)  &  &  & 1.19 & (1.17, 1.21) \\ 
				{Rate Ratios} \\ 
				\hspace{10pt}DES &  &  & 0.87 & (0.82, 0.93)  &  &  & 0.95 & (0.89, 1.01) \\ 
				\hspace{10pt}Mother Smoked &  &  & 0.98 & (0.96, 1.00)   &  &  & 0.98 & (0.96, 1.00) \\ 
				\hspace{10pt}{Year of birth} \\
				\hspace{10pt}\hspace{10pt}1951-1955  &  &  & 1.02 & (1.00, 1.04)   &  &  & 1.02 & (0.99, 1.04) \\ 
				\hspace{10pt}\hspace{10pt}1956-1960  &  &  & 1.08 & (1.05, 1.10)   &  &  & 1.08 & (1.05, 1.10) \\ 
				\hspace{10pt}\hspace{10pt}1961-1965  &  &  & 1.12 & (1.09, 1.16)   &  &  & 1.12 & (1.09, 1.16) \\ 
				\\[-1.8ex]
				{\emph{ADHD}} \\
				{Baseline Odds} & 0.04 & (0.03, 0.04) & 0.03 & (0.03, 0.04) & 0.04 & (0.03, 0.04) & 0.03 & (0.03, 0.04) \\ 
				{Odds Ratios} \\ 
				\hspace{10pt}DES  & 1.39 & (1.13, 1.71) & 1.42 & (1.16, 1.73)   & 1.33 & (1.03, 1.71) & 1.44 & (1.18, 1.75) \\ 
				\hspace{10pt}Mother Smoked  & 1.23 & (1.14, 1.32) & 1.26 & (1.17, 1.35)  & 1.23 & (1.14, 1.32) & 1.26 & (1.17, 1.35) \\ 
				\hspace{10pt}{Year of birth} \\
				\hspace{10pt}\hspace{10pt}1951-1955  & 1.58 & (1.44, 1.73) & 1.62 & (1.48, 1.77)   & 1.58 & (1.43, 1.75) & 1.62 & (1.48, 1.77) \\ 
				\hspace{10pt}\hspace{10pt}1956-1960  & 1.83 & (1.67, 2.01) & 1.84 & (1.68, 2.02)   & 1.83 & (1.66, 2.02) & 1.84 & (1.68, 2.02) \\ 
				\hspace{10pt}\hspace{10pt}1961-1965 & 1.64 & (1.46, 1.85) & 1.65 & (1.47, 1.85)   & 1.64 & (1.46, 1.85) & 1.65 & (1.47, 1.85) \\ 
				$\sigma_0$  &  &  & 2.03 & (1.95, 2.12)    &  &  & 2.04 & (1.95, 2.12) \\ 
				$\sigma_1$   &  &  & 1.69 & (1.18, 2.19)   &  &  & 1.66 & (1.19, 2.12) \\ 
				$\gamma_0$   &  &  & -0.01 & (-0.01, 0.00)    &  &  & -0.01 & (-0.01, 0.00) \\ 
				$\gamma_1$ &  &  & 0.01 & (-0.05, 0.06)    &  &  & 0.00 & (-0.04, 0.05) \\ 
				\bottomrule
		\end{tabular} }
	\end{table}

	Given the relationship between ADHD and family size (see Table \ref{tab:ICS_tab}), ICS was a natural concern, a priori. Ultimately, though, informativeness was quite low, as indicated by the small estimated scaling factors in the JMM ($\hat{\gamma}_0$=$-$0.01, $\hat{\gamma}_1$=0.00). 
	We thus would not expect bias amplification due to induced differential misclassification (as in Section \ref{ss:induced_differential}). Nevertheless, our simulations implied that the observed degree of misclassification---and misclassification-size relationship---could yield bias even in the absence of ICS. This is true only on average, however, and in this case our results largely confirm those of the original study. 

	A feature of the JMM framework is the ability to study exposure-size associations, and here we do in fact observe a difference between na\"ive and corrected estimates. The na\"ive rate ratio estimate for DES and family size was 0.87 (95\% CI [0.82, 0.93])---implying families exposed to DES are smaller---and the corrected estimate was 0.95 (95\% CI [0.89, 1.01]). 
	Such a correction is particularly meaningful given the level of uncertainty: the magnitude of this shift amounts to 70\% of the width of the na\"ive 95\% CI.

	\section{Discussion}
	\label{s:discuss3}
	We investigated misclassification in cluster-level exposures related to both outcomes and cluster size.  As a special case, we found that misclassification that depends on informative cluster size actually induces differential misclassification with respect to outcomes and can generate complex bias structures. We then proposed two correction frameworks: an observed likelihood approach and an expected estimating equations approach. In the motivating multigenerational study of DES and ADHD, misclassification did in fact depend on cluster size---with sensitivity ranging from 70\% in the smallest clusters to 50\% in the largest---but the corrected estimates of the DES-ADHD association ultimately did not differ meaningfully from the na\"ive ones, thus confirming the original study results. By contrast, corrected and na\"ive estimates of the association between DES and family size differed substantially.
	
	A key takeaway is that misclassification in an exposure related to both outcomes and cluster size can actually induce informative cluster size in the na\"ive model that contains the misclassified exposure, even when cluster size is non-informative in the true model. Nevertheless we showed in  supplementary simulations (see supplementary material for results under the strongly induced informativeness of Figure \ref{fig:mis_box}) that standard correction methods can proceed without modelling cluster size---although care should be taken to respect dependence on cluster size in the misclassification and exposure models (for the observed likelihood), and in the conditional exposure model (for EEE). 
	
	When modelling correlated data, analysts choose between marginal and conditional models based on desired parameter interpretations (population-averaged or cluster-specific effects). Under informative cluster size, marginal inference typically proceeds by joint marginalized modelling of cluster size and outcome or by solving weighted estimating equations, and conditional inference typically proceeds via joint models or GLMMs. Though we have focused on marginal models, an important corollary is that the observed likelihood correction applies equally to cluster-specific inference in joint conditional models---in a supplementary simulation, the  observed likelihood correction performed similarly for conditional inference (see supplementary material). Moreover, observed likelihood for joint models suggests a path forward when clusters may be informatively empty \citep{mcgee2019informatively}.
	
	If one were willing to (and able to correctly) specify the joint distribution for outcome and cluster size, the EEE approach guarantees consistent estimation \citep{wang2000expected}. As a pragmatic alternative, we describe a procedure that directly models the conditional exposure distribution, ignoring the full model structure. This approach is hence an approximate method, not unlike regression calibration, a ubiquitous approximate correction for mismeasured continuous exposures \citep{carroll2006measurement}. But whereas regression calibration does not ensure consistency under non-identity link functions (even when the reclassification model is correctly specified), EEE guarantees consistency if one correctly specifies the conditional exposure distribution. In practice, overly simplistic specifications for the conditional exposure model only partially corrected for bias in all but the simplest cases, and we recommend that analysts specify this model as flexibly as is feasible. If one does specify the full joint distribution of size and outcome in (\ref{eqn:decomp}), misspecification is possible---but this misspecification affects only the weights in (\ref{eqn:AEE1b}), not the complete-data scores themselves. An avenue for future research is to examine whether EEE is more robust to this misspecification than the full likelihood approach.

	\backmatter
	
	
	\section*{Acknowledgements}
	This research was supported by grants from the U.S. National Institute of Environmental Health Sciences (U2C ES026555, P30 ES000002, P30 ES009089) and by the Escher Fund for Autism. The Nurses’ Health Study II is also supported by an infrastructure grant from the National Institutes of Health (UM1 CA176726). Data provided by the Channing Division of Network Medicine, Department of Medicine, Brigham and Women's Hospital and Harvard Medical School. \vspace*{-8pt}
	
	
	\section*{Software}
	Software for implementing the methodology considered here, along with code for running the simulation in Section 6 and the analysis in Section 7, is available at\\ https://github.com/glenmcgee/misclassICS \vspace*{-8pt}
	
	\section*{Supporting Information}
	Additional supporting information may be found online in the Supporting Information section at the end of the article.\vspace*{-8pt}

	
	%

	\appendix

	\label{lastpage}
	
\end{document}